\documentclass[twocolumn,showpacs,preprintnumbers,amsmath,amssymb,floatfix,prd]{revtex4}
\usepackage{epsfig}
\usepackage{dcolumn}
\begin{document}
\title{Parton-to-Pion Fragmentation Reloaded}
\author{Daniel de Florian}
\email{deflo@df.uba.ar} 
\affiliation{Departamento de F\'{\i}sica and IFIBA,  Facultad de Ciencias Exactas y Naturales, Universidad de Buenos Aires, Ciudad Universitaria, Pabell\'on\ 1 (1428) Buenos Aires, Argentina}
\author{Manuel Epele}\email{manuepele@gmail.com}
\affiliation{ Instituto de F\'{\i}sica La Plata, CONICET - UNLP,
Departamento de F\'{\i}sica,  Facultad de Ciencias Exactas, Universidad de
La Plata, C.C. 69, La Plata, Argentina}
\author{Roger J.\ Hern\'andez-Pinto}
\email{roger@df.uba.ar}
\affiliation{Departamento de F\'{\i}sica and IFIBA,  Facultad de Ciencias Exactas y Naturales, Universidad de Buenos Aires, Ciudad Universitaria, Pabell\'on\ 1 (1428) Buenos Aires, Argentina}
\affiliation{Instituto de F\'isica Corpuscular, Universitat de Val\`encia-Consejo Superior de Investigaciones Cient\'ificas, Parc Cientific, E-46980 Paterna, Valencia, Spain}
\author{R. Sassot}
\email{sassot@df.uba.ar} 
\affiliation{Departamento de F\'{\i}sica and IFIBA,  Facultad de Ciencias Exactas y Naturales, Universidad de Buenos Aires, Ciudad Universitaria, Pabell\'on\ 1 (1428) Buenos Aires, Argentina}
\author{Marco Stratmann}
\email{marco.stratmann@uni-tuebingen.de}
\affiliation{Institute for Theoretical Physics, University of T\"ubingen, Auf der Morgenstelle 
14, 72076 T\"ubingen, Germany}

\begin{abstract}
We present a new, comprehensive global analysis of parton-to-pion fragmentation functions at next-to-leading order
accuracy in QCD. The obtained results are based on the latest experimental information on single-inclusive pion production in electron-positron annihilation, lepton-nucleon deep-inelastic scattering, and proton-proton collisions.
An excellent description of all data sets is achieved, and the remaining uncertainties in 
parton-to-pion fragmentation functions are estimated based on the Hessian method.
Extensive comparisons to the results from our previous global analysis are performed.
\end{abstract}


\maketitle

\section{Introduction and Motivation}
%
The quantitative description of hard scattering processes 
involving identified light hadrons in the final-state
requires a precise knowledge of how quarks and gluons
hadronize. In the framework of perturbative QCD (pQCD),
which we pursue in the following, this vital information 
is encoded in parton-to-hadron fragmentation functions (FFs) \cite{ref:collins-soper}. 
To match the increasing amount and precision of experimental results,
the availability of reliable sets of FFs for a large variety of hadrons,
in particular, for neutral and charged pions and kaons, 
as well as accurate estimates of their uncertainties is of the utmost relevance
and the subject of this study. 

Like parton distribution functions (PDFs), FFs are 
required in a pQCD calculation to consistently absorb certain 
classes of collinear parton-parton configurations
related to long-distance physics, i.e., interactions
happening a long time after the actual hard scattering process.
As such, FFs are non-perturbative quantities, and any information about them
needs to be gathered from data, preferably in a global QCD analysis
combining results obtained in a large variety of processes.
These fits of FFs are facilitated by assuming factorization \cite{ref:fact}, 
which allows one to compute the relevant 
short-distance hard scattering matrix elements perturbatively, 
and the fact that pQCD predicts the scale evolution of FFs
very much in the same way as for PDFs.
Fragmentation functions depend on the parton of flavor $i$ which hadronizes, 
the fraction $z$ of its four-momentum taken by the observed hadron $H$,
and the scale $Q$ at which they are probed in a hard-scattering process.
In what follows, they will be denoted as $D_i^H(z,Q^2)$. 
All relevant ingredients for a global QCD analysis of FFs are fully known up to 
next-to-leading order (NLO) accuracy in the strong coupling $\alpha_s$ comprising 
the kernels governing the time-like scale evolution \cite{ref:kernel-nlo,ref:sv-kernel},
single-inclusive hadron production in electron-positron annihilation
(SIA) \cite{ref:sia-nlo,ref:sia-neerven-long-as2,ref:lambda-nlo} 
and proton-proton ($pp$) collisions \cite{ref:pp-nlo},
and hadron multiplicities in semi-inclusive deep-inelastic 
lepton-nucleon scattering (SIDIS) \cite{ref:sidis-nlo,ref:lambda-nlo}.

A first global QCD analysis of data collected for all these hadron 
production processes in terms of pion (and kaon) FFs has been performed 
quite some time ago in Ref.~\cite{ref:dss}, commonly known as ``DSS analysis'',
followed by similar fits for eta mesons, protons, and unidentified charged hadrons \cite{ref:dss2,ref:eta}.
Compared to earlier studies based on SIA data \cite{ref:kretzer,ref:other-ffs}, 
the DSS fit \cite{ref:dss} fully exploited the synergy of the complementary probes of FFs 
(SIA, SIDIS, and $pp$) to arrive at a more detailed and data-driven separation of the 
individual FFs for different quark flavors than before.
The use of charge separated, i.e., $H=\pi^{\pm}, K^{\pm}$, SIDIS multiplicities \cite{ref:hermes-old}
was instrumental in this respect as they weight quite differently contributions 
of the various quark flavors $i$ in the hadronization process.
While SIA data are more straightforward to analyze and free of PDF uncertainties,
the LEP \cite{ref:alephdata,ref:delphidata,ref:opaldata}
and SLAC \cite{ref:slddata} data used in the DSS fit 
only provided a handle on flavor-separated FFs when
supplemented by corresponding flavor-tagged results \cite{ref:delphidata,ref:slddata,ref:opaleta},
which have no unambiguous theoretical
description \cite{ref:kretzer,ref:dss} and heavily rely on Monte Carlo simulations  
to extract them experimentally \cite{ref:opaleta}.
One peculiar finding of the DSS global analysis \cite{ref:dss} was an unexpectedly large charge symmetry violation
between the total $u$- and $d$-quark FFs for pions, within sizable uncertainties though.
In addition, single-inclusive pion data from $pp$ collisions at BNL-RHIC \cite{ref:phenixdata} provided a first
constraint on the gluon-to-pion FFs, which, at that time, was impossible to
determine otherwise as precise enough SIA data were only available from the
LEP and SLAC experiments, i.e., at a fixed scale $M_Z$, the mass of the $Z$ boson. 

To implement the lengthy, exact NLO expressions for hadron production processes in SIDIS and $pp$ 
collisions without any approximations into the theoretical framework of a global fit, DSS adopted the 
Mellin technique \cite{ref:mellin1,ref:mellin2}. 
The gist of this method is to pre-calculate all time-consuming NLO expressions for
SIDIS and $pp$ processes once, before the actual fit is performed, and to store the required 
information on look-up tables; for details, see \cite{ref:mellin2,ref:dss,ref:dssv}.
The use of Mellin moments is also most appropriate in solving the QCD evolution equations. 

Uncertainties of the extracted FFs were estimated based on the robust Lagrange multiplier (LM)
technique \cite{ref:lm}, but only for a specific moment of the $D_i^H(z,Q^2)$ contributing to the
momentum sum rule. More detailed studies were performed in Ref.~\cite{ref:dss-unc}, where, in addition,
the applicability of the standard iterative Hessian (IH) approach \cite{ref:ih} was explored. 
Comparisons with the results obtained with the LM technique revealed, however, 
some limitations of the IH method, mainly due to the lack of sufficiently 
precise experimental information at that time
to warrant the assumption that any deviations from the optimum fit are
quadratic in all the parameters specifying the FFs.

In the present paper we build upon the theoretical and conceptual framework developed
for the DSS analysis \cite{ref:dss} but make use of a wealth of newly available data sets,
which will enable us to relax and scrutinize some of the constraints imposed on the parameter 
space in the DSS fit. The key assets of the new analysis are the recently published 
precise SIA data from {\sc BaBar} \cite{ref:babardata} and {\sc Belle} \cite{ref:belledata}, 
which, in principle, should provide a novel handle on the gluon FF through QCD scaling violations 
of the SIA structure functions between the scale $Q=M_Z$, relevant for the LEP and SLAC experiments, 
and the scale corresponding to the center-of-mass system (c.m.s.)
energy of {\sc BaBar} and {\sc Belle}, $Q=\sqrt{S}\simeq 10.5\,\mathrm{GeV}$. 
In addition, since the electroweak couplings of up-type and down-type quarks to the $Z$ boson
become almost equal at $Q \approx M_Z$, LEP and SLAC data are mainly sensitive to the total quark
singlet FF for any observed hadron $H$. At the lower $\sqrt{S}$ of {\sc BaBar} and {\sc Belle}, 
the quark-antiquark pairs in SIA are weighted according to their electrical charge, 
which in our global fit should allow for some partial flavor separation of FFs.

Another important and new ingredient to the current analysis are the final SIDIS data released by
the {\sc Hermes} Collaboration \cite{ref:hermesmult} which supersede the preliminary and 
much less precise data utilized in the DSS fit \cite{ref:hermes-old}.
New, still preliminary data for pion multiplicities in SIDIS are also available, for the first time,
from the {\sc Compass} experiment at CERN \cite{ref:compassmult}, which are very precise 
despite exhibiting a fine binning in the relevant kinematic variables. 
Finally, first results on single-inclusive pion spectra at high transverse momenta $p_T$
have become available from the LHC at c.m.s.\ energies of up to $7\,\mathrm{TeV}$ \cite{ref:alicedata}, which
supplement the data from BNL-RHIC taken at $\sqrt{S}=200\,\mathrm{GeV}$ that have been already used in 
the original DSS analysis \cite{ref:phenixdata}. We also include several recent results from the {\sc Star} Collaboration
for both neutral and charged pion production at $\sqrt{S}=200\,\mathrm{GeV}$ \cite{ref:starcharged06,ref:stardata09,ref:starratio11,ref:stardata13}.
We note that at variance with the original DSS analysis, we now determine the optimum
normalization shifts for each data set in the fit analytically (see, e.g.~Ref.~\cite{ref:lm}
for a discussion of normalization shifts in PDF fits), 
which greatly facilitates the global fitting procedure.

The main goal of our new analysis is to extract an updated set of parton-to-pion FFs and to
determine their uncertainties reliably based on the IH method \cite{ref:ih}
in light of all the newly available, precise experimental results in SIA, SIDIS, and $pp$ collisions.
This will allow us to scrutinize the consistency of the information on FFs extracted across the different
hard scattering processes, i.e., to validate the fundamental notion of universality, 
which is at the heart of any pQCD calculation based on the factorization of short- 
and long-distance physics \cite{ref:fact}
sketched above. 

For the time being, we have to limit ourselves to pion FFs as a similar level of
improvements on the available experimental information is still lacking for kaons,
most noticeable for the SIDIS process, which is crucial in determining flavor-separated FFs.
Nevertheless, we strongly believe that 
our updated global analysis of parton-to-pion FFs is very timely for the
reasons mentioned above and the fact that precise FFs are in high demand as input
for global analyses of helicity PDFs \cite{ref:dssv,ref:pdf-appl} and transverse momentum dependent PDFs \cite{ref:tmd-appl}, 
both of which heavily draw on data with identified pions in the final-state.
Other applications involve to quantify and understand possible modifications of
hadron production yields in the presence of a nuclear medium, as studied in heavy ion
collisions both at RHIC and the LHC \cite{ref:modff}.

Since extractions of leading order (LO) FFs have yielded a much less satisfactory description
of the available pion production data in the DSS analysis \cite{ref:dss}, we only 
perform our global QCD fit at NLO accuracy. In any case, the need for LO FFs (and PDFs) 
has greatly diminished in recent years with the advent of novel
theoretical tools that allow one to compute NLO cross sections 
largely automatically.
The obtained optimum NLO parton-to-pion FFs, including the Hessian eigenvector
sets, are available upon request and enable one to straightforwardly 
propagate our obtained uncertainties to any observable of interest.

It should be noted that the necessary time-like evolution kernels for FFs 
are available even at next-to-NLO (NNLO) accuracy now, with the exception of one,
presumably minor, detail for phenomenological applications
\cite{ref:nnlo-kernel}. However, the corresponding partonic hard scattering 
processes have been only computed for SIA so far \cite{ref:nnlo-sia}. Nevertheless, it might be an interesting
future endeavor to perform a NNLO analysis of SIA data alone and, perhaps, to investigate
the impact of also available all-order resummations of potentially large logarithmic
corrections near the partonic threshold \cite{ref:resum}.
This is, however, well beyond the scope of the current analysis.

The remainder of the paper is organized as follows: in the next Section we briefly summarize
the main aspects of our updated global analysis, including the choice for the 
functional form used to parametrize the FFs at the initial scale for the QCD evolution, 
the selection of data sets  and cuts imposed on them, and the treatment of 
experimental normalization uncertainties.
The outcome of the new fit is discussed in depth in Sec.~III. 
The obtained parton-to-pion fragmentation functions and their uncertainties
are shown and compared to the results of our previous global analysis.
Detailed comparisons to the individual data sets are given 
to demonstrate the quality of the fit. Potential open issues and tensions among
the different data sets will be discussed.
We briefly summarize the main results in Sec.~IV.

\section{Framework}
%
In this Section we lay out the framework and key ingredients for our
global QCD analysis of parton-to-pion FFs. We mainly focus on those
aspects that differ from the original DSS analysis \cite{ref:dss}.
%
\subsection{Functional Form and Fit Parameters \label{sec:funcform}}
%
The functional form adopted in the DSS global analysis \cite{ref:dss}  
is flexible enough to accommodate also the wealth of new experimental 
information included in the present fit.
Therefore, we continue to parametrize the hadronization of a parton of flavor $i$ 
into a positively charged pion at an initial scale of $Q_0=1\,\mathrm{GeV}$ as
\begin{equation}
\label{eq:ff-input}
D_i^{\pi^+}\!(z,Q_0) =
\frac{N_i z^{\alpha_i}(1-z)^{\beta_i} [1+\gamma_i (1-z)^{\delta_i}] }
{B[2+\alpha_i,\beta_i+1]+\gamma_i B[2+\alpha_i,\beta_i+\delta_i+1]}\;.
\end{equation}
Here, $B[a,b]$ denotes the Euler Beta-function, and the 
$N_i$ in (\ref{eq:ff-input}) are chosen in such a way that
they represent the contribution of $z D_i^{\pi^+}$ to the momentum
sum rule.

Compared to our previous analysis, the improved experimental information
now allows us to impose less constraints on the parameter space spanned
by the input function in Eq.~(\ref{eq:ff-input}). More specifically,
as before we still have to assume isospin symmetry for the unfavored FFs
of light sea quarks, i.e.,
\begin{equation}
\label{eq:su2constraint}
D_{\bar{u}}^{\pi^+}=D_{d}^{\pi^+}\;,
\end{equation}
and we need to relate the total $u$-quark and $d$-quark FFs 
by a global, $z$-independent factor $N_{d+\bar{d}}$,
\begin{equation}
\label{eq:su2breaking}
D_{d+\bar{d}}^{\pi^+} = N_{d+\bar{d}} D_{u+\bar{u}}^{\pi^+},
\end{equation}
which quantifies any charge symmetry violation found in the fit.
The fragmentation of a strange quark into a pion
is now related to the unfavored FFs 
in Eq.~(\ref{eq:su2constraint}) by 
\begin{equation}
\label{eq:su3breaking}
D_s^{\pi^+}=D_{\bar{s}}^{\pi^+} = N_s z^{\alpha_s} D_{\bar{u}}^{\pi^+}
\end{equation}
rather than just using a constant as in the DSS analysis.

The charm- and bottom-to-pion FFs no longer assume 
$\gamma_c=\gamma_b=0$ in Eq.~(\ref{eq:ff-input}) but can now exploit the 
full flexibility of the ansatz. This is not due to new flavor-tagged data
but helps the global fit to accommodate the recent, very precise results from
{\sc BaBar} \cite{ref:babardata} and {\sc Belle} \cite{ref:belledata} in SIA 
and from {\sc Compass} \cite{ref:compassmult} and {\sc Hermes} \cite{ref:hermesmult}
in SIDIS, which now constrain both the total quark fragmentation, i.e.,
summed over all flavors, and the individual flavor-separated, light quark 
FFs much better than before. As in the DSS and all 
other analyses \cite{ref:dss,ref:dss2,ref:eta,ref:kretzer,ref:other-ffs}, we include heavy flavor FFs 
discontinuously as massless partons in the QCD scale evolution 
above their $\overline{\text{MS}}$ ``thresholds'', $Q=m_{c,b}$,
with $m_c$ and $m_b$ denoting the mass of the charm and bottom quark,
respectively. 
Conceptually, due to confinement, there has to be a heavy quark FFs present as
soon as the heavy quark can be produced in the final-state of a
hard scattering process. We leave it to dedicated future studies to
explore and incorporate an improved theoretical framework for heavy quark-to-light hadron
fragmentation functions into the global fitting procedure, following the
rather elaborate schemes that have been developed for
heavy quark parton densities \cite{ref:hq-pdfs} to properly include mass effects near threshold 
and to resum potentially large logarithms $\sim \ln m_{c,b}^2/Q^2$ for $Q^2\gg m_{c,b}^2$.
We note that a dynamical, parameter-free generation of the heavy flavor component to light
meson fragmentation functions has been developed, for instance, in Ref.~\cite{ref:hq-ffs}.

In total we now have 28 free fit parameters describing our updated FFs for quarks,
antiquarks, and gluons into positively charged pions, which are 
determined from data by a standard $\chi^2$ minimization to be described below.
The corresponding FFs for negatively charged pions are obtained by charge conjugation 
and those for neutral pions by assuming $D_i^{\pi^0}= [D_i^{\pi^+}+D_i^{\pi^-}]/2$.
We note that none of the constraints imposed on the fit
through Eqs.~(\ref{eq:su2constraint})-({\ref{eq:su3breaking})
has any impact on its overall quality. 

\subsection{Data Selection \label{sec:datasets}}
%
We make use of all the currently available experimental information on single-inclusive
charged and neutral pion production in SIA, SIDIS, and hadron-hadron collisions to
determine the free fit parameters defined in Sec.~\ref{sec:funcform}.

Compared to the data sets already used in the DSS global analysis \cite{ref:dss}, we include the new
results from {\sc BaBar} \cite{ref:babardata} and {\sc Belle} \cite{ref:belledata} in SIA 
at a c.m.s.\ energy of $\sqrt{S}\simeq 10.5\,\mathrm{GeV}$.
Both sets are very precise, with relative uncertainties of about $2-3\%$, and reach
all the way up to pion momentum fractions $z$ close to one, well beyond of what has been measured so far.
We analyze both sets with $n_f=4$ active, massless flavors using the standard expression for 
the NLO SIA cross section \cite{ref:sia-nlo}. As customary, we
limit ourselves to data with $z\ge 0.1$ to avoid any potential impact from kinematical
regions where finite, but neglected, hadron mass corrections,
proportional to $M_{\pi}/(S z^2)$, might become of any importance \cite{ref:dss,ref:kretzer,ref:other-ffs}. 
For SIA data taken at higher $\sqrt{S}$ we use $n_f=5$ and $z>0.05$, following the original DSS analysis.
Any incompatibility of the two new precise sets of data at $\sqrt{S}\simeq 10.5\,\mathrm{GeV}$ 
with each other or with the old LEP and SLAC data at $\sqrt{S}\simeq 91.2\,\mathrm{GeV}$ 
\cite{ref:alephdata,ref:delphidata,ref:opaldata,ref:slddata}
has the potential to seriously spoil the quality of the global fit.

In case of SIDIS, we replace the preliminary multiplicity data from {\sc Hermes} \cite{ref:hermes-old}
by their recently released final results \cite{ref:hermesmult}.
More specifically, we use the data for charged pion multiplicities as a function of momentum transfer $Q^2$ 
in four bins of $z$ taken on both a proton and a deuteron target. 
The range of average values of $Q^2$ 
covered by the data is from about $1.1\,\mathrm{GeV}^2$ to $7.4\,\mathrm{GeV}^2$ and $0.2\le z \le 0.8$.
In addition, we include the still preliminary multiplicity data for $\pi^{\pm}$ 
from the {\sc Compass} Collaboration \cite{ref:compassmult}, which are given as a function of $z$ in bins 
of $Q^2$ and the initial-state momentum fraction $x$. 
The coverage in $z$ is the same as for the {\sc Hermes} data, but due to the higher $\sqrt{S}$ of
the {\sc Compass} experiment the reach in $x$ and $Q^2$ is wider. Experimental information is
available for $0.004\le x \le 0.7$ and $1.2 \le Q^2 \le 22.4\,\mathrm{GeV}^2$.
We do not have to impose any cuts on both data sets to accommodate them in the global analysis.
As for SIA, having now available two precise sets of multiplicity data in SIDIS, covering slightly 
different but partially overlapping kinematics, makes it
very important to validate their consistency in a global fit.

Finally, we add a couple of new sets of data for inclusive high-$p_T$ pion production in $pp$ collisions 
to the results from the {\sc Phenix} experiment \cite{ref:phenixdata} already included in the DSS analysis. 
Most noteworthy are the first results for neutral pions from the {\sc Alice} Collaboration 
at CERN-LHC \cite{ref:alicedata}, covering 
unprecedented c.m.s.\ energies of up to $7\,\mathrm{TeV}$. In addition, we
add {\sc Star} data taken at $\sqrt{S}=200\,\mathrm{GeV}$ in various rapidity intervals
for both neutral and charged pion production and for the $\pi^-/\pi^+$ ratio
\cite{ref:starcharged06,ref:stardata09,ref:starratio11,ref:stardata13}. 
As we will demonstrate and discuss in more detail in Sec.~\ref{sec:pp-results} below, 
it turns out that a good global fit of RHIC and LHC $pp$ data, along with all the other world data, 
can only be achieved if one imposes a cut on the 
minimum $p_T$ of the produced pion of about $5\,\mathrm{GeV}$. Such a cut eliminates some
of the $pp$ data points included in the previous DSS analysis from the fit, in particular, 
all the {\sc Brahms} \cite{ref:brahmsdata} and {\sc Star} \cite{ref:starforward}
data at forward pseudo-rapidities and, hence, too small values of $p_T$.

\subsection{Fit Procedure and Uncertainty Estimates \label{sec:uncertainties}}
%
The 28 free parameters describing the updated parton-to-pion FFs 
in Eq.~(\ref{eq:ff-input}) at the chosen input scale of $1\,\mathrm{GeV}$ are again
determined from a standard $\chi^2$ minimization where
\begin{equation}
\label{eq:chi2}
\chi^2=\sum_{i=1}^N  \left[
\left( \frac{1-{\cal{N}}_i}{\delta{\cal{N}}_i}  \right)^2 +
\sum_{j=1}^{N_i}
\frac{({\cal{N}}_i T_j-E_j)^2}{\delta E_j^2} \right],
\end{equation}
for $i=1,\ldots,N$ data sets, each contributing with $N_i$ data points.
$E_j$ is the measured value of a given observable,
$\delta E_j$ the error associated with this measurement, and
$T_j$ is the corresponding theoretical estimate for a
given set of parameters in Eq.~(\ref{eq:ff-input}).
Since the full error correlation matrices are not available for
some of the data sets used in the fit, statistical and systematical errors 
are simply added in quadrature in $\delta E_j$
as in all previous fits \cite{ref:dss,ref:dss2,ref:kretzer,ref:other-ffs}.

At variance with the original DSS fit \cite{ref:dss}, where we have introduced several extra fit
parameters to account for experimental normalization uncertainties ${\cal{N}}_i$ in (\ref{eq:chi2}),
we now derive the optimum normalization shifts for each data set
analytically from the condition $\partial \chi^2/\partial {\cal{N}}_i=0$,
which yields
\begin{equation}
\label{eq:normshift}
{\cal{N}}_i = \frac{ \sum_{j=1}^{N_i} \frac{\delta{\cal{N}}_i^2}{\delta E_j^2} T_j E_j +1}
{1+ \sum_{j=1}^{N_i}  \frac{\delta{\cal{N}}_i^2}{\delta E_j^2} T_j^2  }\;.
\end{equation}
Here, $\delta {\cal{N}}_i$ denotes the quoted experimental normalization uncertainty
for data set $i$.
In Sec.~\ref{sec:ff-results} we will list the so obtained normalizations ${\cal{N}}_i$
along with the individual $\chi^2$ values for each data set included in the
fit.

In the DSS analysis \cite{ref:dss} we assessed uncertainties in the extraction of fragmentation functions
with the help of the LM technique \cite{ref:lm} by mapping
out the maximum allowed range of variation in the fit of the truncated second
moments of the fragmentation functions
\begin{equation}
\label{eq:truncmom}
\eta^{\pi^{+}}_i(x_{\min},Q^2) \equiv \int_{x_{\min}}^1 z D_i^{\pi^{+}}(z,Q^2) dz, 
\end{equation}
for $x_{\min}=0.2$ and $Q=5\,\text{GeV}$.
While this method is very robust, even when some of the fit parameters are only loosely constrained by data,
it has the disadvantage that uncertainties cannot be easily propagated to other observables of interest.
In Ref.~\cite{ref:dss-unc} we have therefore explored the applicability of the iterative Hessian
approach \cite{ref:ih} based on the original DSS choice of data sets, cuts, and parameters
by comparing its outcome to uncertainty estimates obtained with the LM method. 
The main idea of the IH method is to assume a quadratic behavior of the $\chi^2$ hyper-surface
of parameter displacements and to express the $\chi^2$ increment from its minimum value 
in terms of combinations of fit parameters that maximize the variation. 
Such an eigenvector representation of the Hessian matrix proves to be extremely suitable
to compute the propagation of uncertainties to arbitrary observables in terms
of a limited number of pre-calculated sets of FF functions 
(in fact, twice the amount of fit parameters). These sets
correspond to fixed displacements along the eigenvector directions of the Hessian matrix. 

With the much increased availability of precise data for the current analysis, 
the sole use of the computationally less demanding IH method to quantify uncertainties of FFs
becomes viable and will be pursued in the following. The obtained eigenvector
sets of FFs will be made available upon request from the authors along with a
parametrization of the optimum fit.
To define the eigenvector sets one has to choose a tolerance parameter 
$\Delta \chi^2$ for the increment in $\chi^2$ which is still
acceptable in the global fit.
Here we proceed as follows: the tolerances for the eigenvector sets
corresponding to $68\%$ and $90\%$ confidence level (C.L.) intervals
are determined from the Gaussian probability density function for a
$\chi^2$ distribution with $k$ degrees of freedom (d.o.f.):
\begin{equation}
P_k(x) = \frac{x^{k/2-1} e^{-x/2}}{\Gamma(k/2)2^{k/2}}\;.
\end{equation}
The $\Delta \chi^2$ related to the $68^{th}$ and $90^{th}$ percentiles are 
then obtained by solving
$\int_0^{\chi^2+\Delta \chi^2} d\chi^2 P_k(\chi^2) = 0.68$ and $0.90$, respectively.

Finally, we choose the NLO set of PDFs from the MSTW group \cite{ref:mstw} and the
corresponding uncertainty estimates in computations of the SIDIS and $pp$ cross sections.
For consistency, we also fix the strong coupling $\alpha_s$ to the values obtained in 
the MSTW fit.
We note that in the $x$ and $Q^2$ region relevant for our global analysis of FFs,
the needed combinations of PDFs are relatively well constrained. A choice of PDFs other
than the MSTW set would not alter the outcome of our fit in any significant way.
We will illustrate in Sec.~\ref{sec:pp-results} below that theoretical scale ambiguities 
are considerably larger than PDF uncertainties.

\section{Results}
%
In this Section we present and discuss in depth the results of our global
analysis of parton-to-pion FFs. 
First, we present the obtained fit parameters,
normalization shifts, and individual $\chi^2$ values. 
Next, the obtained $D_i^{\pi^{+}}(z,Q^2)$ and their uncertainties
are shown and compared to the results of the DSS fit.
The quality of the fit to SIA, SIDIS, and $pp$ data and potential
open issues and tensions among the different sets of data
are illustrated and discussed in Sec.~\ref{sec:sia-data}, \ref{sec:sidis-data},
and \ref{sec:pp-results}, respectively.
%
\subsection{Parton-To-Pion Fragmentation Functions \label{sec:ff-results}}
%
%
\begin{table}[th!]
\caption{\label{tab:nlopionpara}Parameters describing the NLO FFs for positively charged
pions, $D_i^{\pi^+}(z,Q_0)$,
in Eq.~(\ref{eq:ff-input}) in the $\overline{\mathrm{MS}}$ scheme at the input scale $Q_0=1\,\mathrm{GeV}$.
Results for the charm and bottom FFs refer to
$Q_0=m_c=1.43\,\mathrm{GeV}$ and
$Q_0=m_b=4.3\,\mathrm{GeV}$, respectively.}
\begin{ruledtabular}
\begin{tabular}{cccccc}
flavor $i$ &$N_i$ & $\alpha_i$ & $\beta_i$ &$\gamma_i$ &$\delta_i$\\
\hline
%
%
$u+\overline{u}$ & 0.4465&-0.455& 0.912& 8.00& 4.14\\
$d+\overline{d}$ & 0.4471&-0.455& 0.912& 8.00& 4.14\\
$\overline{u}=d$ & 0.127& 0.997& 2.884&31.48& 7.70\\
$s+\overline{s}$ & 0.378& 0.546& 2.884&31.48& 7.70\\
$c+\overline{c}$ & 0.348& 0.555& 4.883& 7.62& 7.64\\
$b+\overline{b}$ & 0.401&-0.262& 4.369&20.80&11.72\\
$g$              & 0.216& 1.542& 4.066&86.34&19.32\\
\end{tabular}
\end{ruledtabular}
\end{table}
In Table~\ref{tab:nlopionpara} we list the obtained set of parameters
specifying our updated, optimum parton-to-pion fragmentation functions at NLO accuracy
at the input scale $Q_0=1\,\text{GeV}$ for the light quark flavors and the gluon,
and at their respective thresholds $Q_0=m_{c,b}$ for the charm and bottom quarks.

Table~\ref{tab:nlopionpara} reveals already a notable difference to one of the findings of the DSS analysis
which preferred an unexpectedly sizable breaking of the charge symmetry between
$u+\bar{u}$ and $d+\bar{d}$ FFs of about $10\%$ \cite{ref:dss}, within large uncertainties though. 
This was mainly driven by the preliminary $\pi^{\pm}$ multiplicities from {\sc Hermes} \cite{ref:hermes-old}
used in the fit at that time. 
Now, with much improved experimental information on charged pion multiplicities
both from {\sc Hermes} \cite{ref:hermesmult} and {\sc Compass} \cite{ref:compassmult} 
and new data on the ratio $\pi^-/\pi^+$ in $pp$ collisions from {\sc Star} \cite{ref:starratio11}, 
the parameter $N_{d+\bar{d}}$ in Eq.~(\ref{eq:su2breaking}) prefers to stay
very close to unity, i.e., very little or no breaking. 

As has been mentioned above, in case of the unfavored FFs, 
data now allow us to introduce some non-trivial $z$ dependence, see Eq.~(\ref{eq:su3breaking}),
to parametrize a potential SU(3) breaking between the $\bar{u}=d$ and
$s=\bar{s}$ FFs. This little extra freedom not only helps to accommodate all the different data
sets used in the global analysis in a much better way, but it also leads to more realistic uncertainty
estimates for both $D_s^{\pi^{+}}$ and $D_{\bar{u}}^{\pi^{+}}$.
In particular, it now turns out that the uncertainties for $D_s^{\pi^{+}}$
are much bigger than for $D_{\bar{u}}^{\pi^{+}}$ as can be inferred from Fig.~\ref{fig:ff-at-10},
where we present the individual parton-to-pion FFs $D_i^{\pi^{+}}(z,Q^2)$ at $Q^2=10\,\mathrm{GeV}^2$.
%
\begin{figure*}[hbt!]
\vspace*{-0.4cm}
\begin{center}
\epsfig{figure=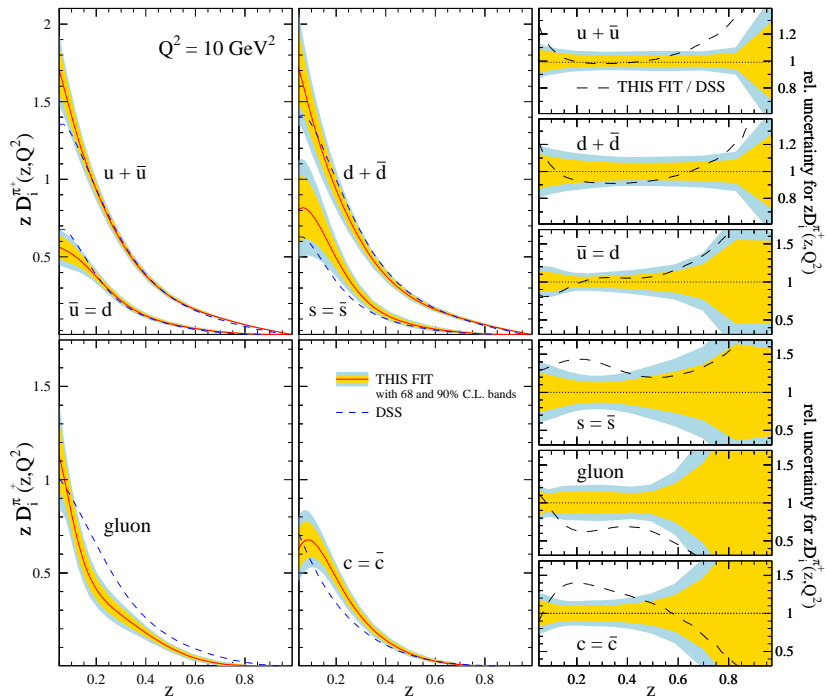,width=0.61\textwidth}
\end{center}
\vspace*{-0.5cm}
\caption{The individual FFs for positively charged pions $zD_i^{\pi^{+}}(z,Q^2)$ at
$Q^2=10\,\mathrm{GeV}^2$ along with uncertainty estimates at $68\%$ and $90\%$ C.L.\
indicated by the inner and outer shaded bands, respectively.
The panels on the right-hand-side show the corresponding relative uncertainties.
Also shown is a comparison to the previous global analysis by DSS \cite{ref:dss} (dashed lines).
\label{fig:ff-at-10}}
\end{figure*}
The four leftmost panels show the optimum $zD_i^{\pi^{+}}$ at NLO accuracy for 
$i=u+\bar{u}$, $d+\bar{d}$, $\bar{u}=d$, $s=\bar{s}$, $c=\bar{c}$, and the gluon $g$
(solid lines) along with our uncertainty estimates at $68\%$ C.L.\ (inner bands) and $90\%$ C.L.\
(outer bands), obtained as described in Sec.~\ref{sec:uncertainties}.
For better visibility, the rightmost panels give the relative uncertainties for the same set of
$zD_i^{\pi^{+}}$. The results of the previous NLO DSS fit are shown as dashed lines. 

As can be inferred from Fig.~\ref{fig:ff-at-10}, for the light quark flavors the old DSS results are either close to
the updated fit or within its $90\%$ C.L.\ uncertainty band.
The best determined pion FFs is $D_{u+\bar{u}}^{\pi^{+}}$, where the relative uncertainties
are below $10\%$ at $90\%$ C.L.\ throughout most of the relevant $z$ range. 
Only for $z\gtrsim 0.8$ the errors rapidly
increase because of the lack of experimental constraints in this region.
The corresponding uncertainties for $D_{d+\bar{d}}^{\pi^{+}}$ turn out to be slightly larger
as they also include possible violations of SU(2) charge symmetry through Eq.~(\ref{eq:su2breaking}).
We stress again, that at variance with the DSS analysis \cite{ref:dss}, the new fit does not favor any
SU(2) breaking. For the unfavored FFs, $D_{\bar{u}}^{\pi^{+}}=D_{d}^{\pi^{+}}$
are determined well in a much more limited range of $z$, and uncertainties start to increase
already for $z\gtrsim 0.5$. The corresponding ambiguities on 
$D_s^{\pi^{+}}=D_{\bar{s}}^{\pi^{+}}$ are about a factor of two larger and amount to at least
$25\%$ at $90\%$ C.L.\ for $z\simeq 0.3$.

Bigger deviations from the DSS analysis are found for both the gluon and the charm FFs.
In the latter case, this is driven by the greater flexibility of the functional form,
five fit parameters rather than three, which helps with the overall quality of the
global fit and cannot be pin-pointed to a particular data set. In fact, there had been 
no new charm (or bottom) tagged data since the LEP and SLAC era.
The significantly reduced $D_g^{\pi^{+}}$ as compared to the DSS fit is a result of the
new {\sc Alice} $pp$ data \cite{ref:alicedata}, which have a strong preference for less pions from
gluon fragmentation for basically all values of $z$.
We will discuss this finding, and possible tensions arising with the $pp$ data from RHIC, 
in more detail in Sec.~\ref{sec:pp-results}.
The relative uncertainties on $D_g^{\pi^{+}}$ at $Q^2=10\,\mathrm{GeV}^2$
are about $20\%$ at $90\%$ C.L.\ up to $z\simeq 0.5$ and quickly increase towards larger $z$ values.

\begin{figure*}[ht!]
\vspace*{-0.4cm}
\begin{center}
\epsfig{figure=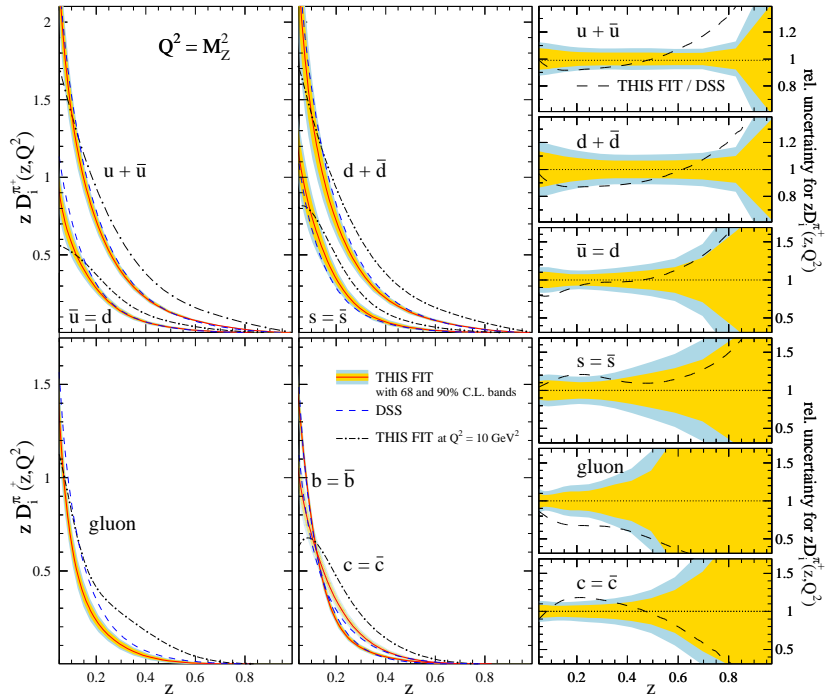,width=0.61\textwidth}
\end{center}
\vspace*{-0.5cm}
\caption{As in Fig.~\ref{fig:ff-at-10} but now for $Q^2=M_Z^2$. Note that
here also a bottom-to-$\pi^+$ fragmentation function is shown.
\label{fig:ff-at-mz}}
\end{figure*}
We refrain from performing a detailed comparison to the uncertainty estimates 
based on the data sets available for the original DSS analysis \cite{ref:dss,ref:dss-unc}
as they can be viewed at best as a rough approximation.
Only with the quality and variety of data sets available for the current
global analysis one can arrive at a first meaningful determination of uncertainties
for parton-to-pion FFs, which therefore constitutes as one of the main results of this study.

We note that the new very precise SIA data from {\sc BaBar} \cite{ref:babardata} and {\sc Belle} \cite{ref:belledata}
help to reliably constrain light quark FFs to much higher values of $z$ than before, in particular,
$D_{u+\bar{u}}^{\pi^{+}}$. 
In combination with the LEP and SLAC data, which, at $Q^2=M_Z^2$, mainly constrain 
the total quark singlet fragmentation function, the new precise data at $\sqrt{S}\simeq 10.5\,\mathrm{GeV}$
also help to provide some partial flavor separation from SIA data alone, as they are sensitive to the
electrical charge weighted sum of quark FFs. Multiplicities in SIDIS for identified charged pions
provide further invaluable experimental input to address this question, see Sec.~\ref{sec:sidis-data} below.
In case of $D_g^{\pi^{+}}$, for the first time, some constraint can be derived from QCD scaling violations in SIA
thanks to having now available two precise sets of data at different energy scales $Q^2\approx 110 \,\mathrm{GeV}^2$
and $Q^2=M_Z^2$. However, scaling violations for FFs in the relevant medium-to-large $z$ range are fairly mild
and also the coverage of the LEP and SLAC data is much more sparse towards high $z$, which, to some extent,
is reflected in the still relatively large uncertainties obtained for $D_g^{\pi^{+}}$ in Fig.~\ref{fig:ff-at-10}.

To demonstrate the scale evolution of FFs, we show in Fig.~\ref{fig:ff-at-mz}
the same $zD_i^{\pi^{+}}$ as in Fig.~\ref{fig:ff-at-10} but now at $Q^2=M_Z^2$.
Since we are above the bottom threshold $Q=m_b$, we now include also our results for
$zD_b^{\pi^{+}}=zD_{\bar{b}}^{\pi^{+}}$ in the middle panel of the lower row.
To facilitate the comparison of the FFs computed at the two different scales, the
dot-dashed lines in Fig.~\ref{fig:ff-at-mz} repeat the results for the new, optimum
fit at $Q^2=10\,\mathrm{GeV}^2$ shown in Fig.~\ref{fig:ff-at-10}. The FFs of the DSS fit are 
again denoted as dashed lines. As can be seen, evolution to larger $Q^2$
reduces the FFs for essentially all relevant $z$ values above about $z\simeq 0.15$.
This trend is reminiscent of the $Q^2$ evolution of PDFs at not to small values
of $x$, which is not surprising as the LO evolution kernels are essentially the same
for the time-like and space-like case.
The increase of the FFs at small $z$ is phenomenologically not relevant as their
range of applicability is anyhow restricted to $z\gtrsim 0.05$.
The relative uncertainties, again given in the rightmost panels of Fig.~\ref{fig:ff-at-mz},
are largely similar to those obtained at $Q^2=10\,\mathrm{GeV}^2$. 
Some of the sizable ambiguities at large $z$ are pushed towards smaller $z$ by evolution,
most noticeable for $D_g^{\pi^{+}}$.

\begin{table}[bth!]
\caption{\label{tab:exppiontab}Data sets used in our NLO global analysis, 
their optimum normalization shifts $N_i$, cf.\ Sec.~\ref{sec:uncertainties} and Eq.~(\ref{eq:normshift}),
the individual $\chi^2$ values 
(including the $\chi^2$ penalty from the obtained $N_i$), 
and the total $\chi^2$ of the fit.}
\begin{ruledtabular}
\begin{tabular}{lcccc}
%
%
experiment& data & norm.  & \# data & $\chi^2$ \\
          & type & $N_i$ & in fit     &         \\\hline
{\sc Tpc} \cite{ref:tpcdata}  & incl.\            &  1.043 & 17 & 17.3 \\
                              & $uds$ tag         &  1.043 &  9 & 2.1 \\
                              & $c$ tag           &  1.043 &  9 & 5.9 \\
                              & $b$ tag           &  1.043 &  9 & 9.2 \\
{\sc Tasso} \cite{ref:tassodata}\hfill 34~GeV &   incl.    & 1.043     & 11 &  30.2    \\
                   \hfill 44~GeV              &   incl.    & 1.043     &  7 &  22.2    \\
{\sc Sld} \cite{ref:slddata}  & incl.\  &  0.986 & 28 & 15.3 \\
          & $uds$ tag         &  0.986 & 17 & 18.5 \\
          & $c$ tag           &  0.986 & 17 & 16.1  \\
          & $b$ tag           &  0.986 & 17 & 5.8 \\
{\sc Aleph} \cite{ref:alephdata}    & incl.\  & 1.020 & 22 &  22.9 \\
{\sc Delphi} \cite{ref:delphidata}  & incl.\  & 1.000  & 17 & 28.3 \\
                     & $uds$ tag   &  1.000  & 17 & 33.3 \\
                     & $b$ tag     &  1.000  & 17 & 10.6 \\
{\sc Opal} \cite{ref:opaldata,ref:opaleta}  & incl.\ & 1.000 & 21 & 14.0 \\
          & $u$ tag &  0.786  & 5 & 31.6 \\
          & $d$ tag &  0.786  & 5 & 33.0  \\
          & $s$ tag &  0.786  & 5 & 51.3 \\
          & $c$ tag &  0.786  & 5 & 30.4 \\
          & $b$ tag &  0.786  & 5 & 14.6 \\
{\sc BaBar} \cite{ref:babardata}     & incl.\ &   1.031 & 45  & 46.4 \\ 
{\sc Belle} \cite{ref:belledata}     & incl.\ &   1.044 & 78  & 44.0 \\    \hline 
{\sc Hermes} \cite{ref:hermesmult}  & $\pi^+$ (p)&  0.980 & 32 & 27.8\\
                               & $\pi^-$ (p)&  0.980 & 32 & 47.8 \\
                               & $\pi^+$ (d)&  0.981 & 32 & 40.3 \\
                               & $\pi^-$ (d)&  0.981 & 32 & 59.1 \\
{\sc Compass} \cite{ref:compassmult} prel.& $\pi^+$ (d) &  0.946     & 199   &  174.2     \\
                               & $\pi^-$ (d)&  0.946     & 199   &  229.0     \\  \hline                               
{\sc Phenix} \cite{ref:phenixdata}   & $\pi^0$ &  1.112 & 15 & 15.8 \\
{\sc Star} \cite{ref:stardata09,ref:stardata13,ref:starcharged06,ref:starratio11} \hfill 0$\le\eta\le 1$   
                                             & $\pi^{0}$                        & 1.161 &  7  & 5.7        \\    
     \hfill $0.8\le\eta\le 2.0$              & $\pi^{0}$                        & 0.954 &  7  & 2.7        \\    
     \hfill $|\eta|<0.5$                     & $\pi^{\pm}$                      & 1.071 &  8  & 4.3        \\    
     \hfill $|\eta|<0.5$                     & $\pi^+$,$\pi^-/\pi^+$            & 1.006 &  16 & 17.2       \\    
{\sc Alice} \cite{ref:alicedata} \hfill 7~TeV& $\pi^0$                          & 0.766 &  11 & 27.7        \\ \hline\hline
{\bf TOTAL:} & & & 973 & 1154.6\\
\end{tabular}
\end{ruledtabular}
\end{table}
The overall quality of the fit is summarized in Tab.~\ref{tab:exppiontab} 
where we list all data sets included in
our global analysis, as discussed in Sec.~\ref{sec:datasets},
along with their individual $\chi^2$ values and the analytically determined
normalization shifts according to Eq.~(\ref{eq:normshift}).
We note that the quoted $\chi^2$ values are based only on fitted data points,
i.e., after applying the cuts mentioned in Sec.~\ref{sec:datasets}, and include
the $\chi^2$ penalty from the $N_i$, i.e., the first term in Eq.~(\ref{eq:chi2}).

Firstly, it is worth mentioning that there is a more than twofold increase in the
number of available data points as compared to the original DSS analysis \cite{ref:dss}. 
Secondly, the quality of the global fit has improved dramatically 
from $\chi^2/{\mathrm{d.o.f.}}\simeq 2.2$ for DSS, see Tab.~II in Ref.~\cite{ref:dss}, 
to $\chi^2/{\mathrm{d.o.f.}}\simeq 1.2$ for the current fit. 
A more detailed comparison reveals that the individual $\chi^2$ values for
the SIA data \cite{ref:alephdata,ref:delphidata,ref:opaldata,ref:slddata,ref:tpcdata,ref:tassodata},
which were already included in the DSS fit, have, by and large,
not changed significantly. 
The description of the fully flavor separated data from {\sc Opal} \cite{ref:opaleta}
in the fit favors a rather large normalization shift but has
nevertheless deteriorated. Given that this set has only 25 data points, 
it is the biggest contributor to the total $\chi^2$. However, in general, flavor-tagged results 
should not be taken too literally as they lack a proper interpretation 
and theoretical framework beyond the lowest order as was already pointed out, e.g., in
Refs.~\cite{ref:kretzer,ref:dss}.

The biggest improvement concerns the SIDIS multiplicities from {\sc Hermes} 
which, in their recently published version \cite{ref:hermesmult}, are now described very well by the 
updated fit. Also, the preliminary charged pion multiplicities from {\sc Compass} \cite{ref:compassmult}
and the new SIA data from {\sc BaBar} \cite{ref:babardata} and {\sc Belle} \cite{ref:belledata}
integrate nicely into the global analysis of parton-to-pion FFs.

Finally, and as we will illustrate in detail in Sec.~\ref{sec:pp-results} below, there is some tension among the
$pp$ data sets from RHIC and the LHC, which forced us to introduce a cut $p_T>5\, \mathrm{GeV}$ 
on the pion's transverse momentum in the current fit to accommodate both of them. 
The obtained individual $\chi^2$ values are
all reasonable, as can be inferred from Tab.~\ref{tab:exppiontab}, 
with the new {\sc Alice} data \cite{ref:alicedata} being on the high side, which largely stems from the
penalty for the still sizable normalization shift. This large shift reflects the preference
of the new {\sc Alice} data for a smaller gluon-to-pion FF than extracted by the original DSS fit
based on RHIC {\sc Phenix} data \cite{ref:phenixdata} alone.
As a result of the $p_T$ cut, the number of $pp$ data in the fit for RHIC has decreased as compared to
the DSS analysis. Both the {\sc Brahms} \cite{ref:brahmsdata} and {\sc Star} \cite{ref:starforward} 
results at forward pseudo-rapidities do not pass the $p_T$ cut anymore, and, hence, are excluded from 
the updated fit. Likewise, we do not consider the {\sc Alice} data taken in $pp$ collisions at $\sqrt{s}=900\,\mathrm{GeV}$
\cite{ref:alicedata}, where only a single point would survive the cut in $p_T$.

\subsection{Electron-Position Annihilation Data\label{sec:sia-data}}
%
In Figs.~\ref{fig:ee-untagged} and \ref{fig:ee-babarbelle} we present
a detailed comparison of the results of our fit and its uncertainties 
at both $68\%$ and $90\%$ C.L.\ with the SIA data already included 
and newly added to the original DSS analysis \cite{ref:dss}, respectively.
In general, the agreement of the fit with SIA data is excellent in the entire energy
and $z$ range covered by the experiments.
For $Q^2=M_Z^2$, the {\sc Delphi} data \cite{ref:delphidata} exhibit some mild tension 
with other sets at the same c.m.s.\ energy in the largest $z$ bins, resulting in a somewhat higher 
individual $\chi^2$ value, as can be gathered from Tab.~\ref{tab:exppiontab}.

%
\begin{figure*}[bht!]
\begin{center}
\epsfig{figure=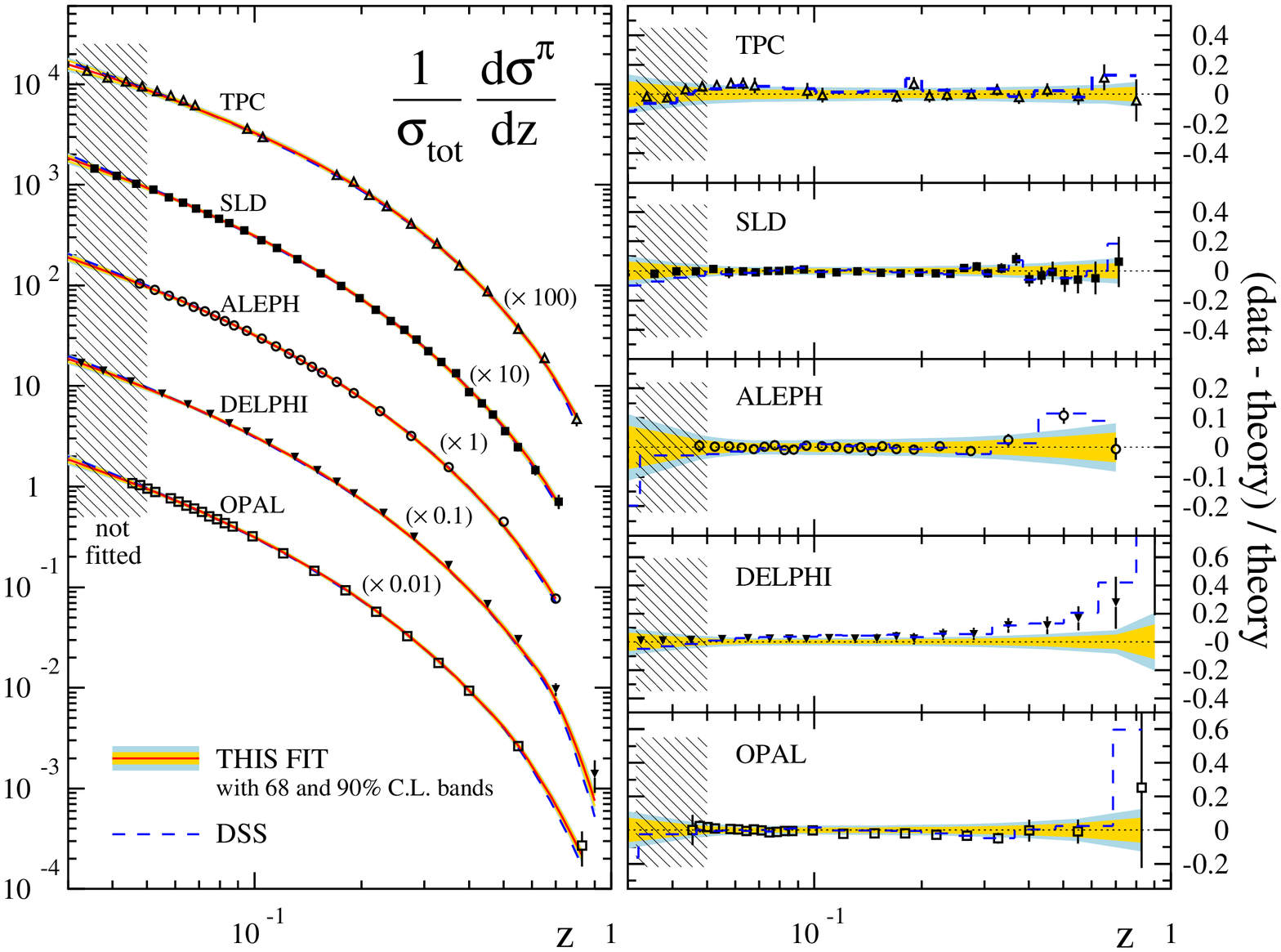,width=0.62\textwidth}
\end{center}
\vspace*{-0.5cm}
\caption{Left-hand side: comparison of our new NLO results (solid lines) and the previous DSS fit \cite{ref:dss} 
(dashed lines)
with data sets for inclusive pion production in SIA used in both fits, see Tab.~\ref{tab:exppiontab}.
The inner and outer shaded bands correspond to new uncertainty estimates at $68\%$ and $90\%$ C.L.,
respectively.
Right-hand side: ``(data-theory)/theory'' for each of the data sets w.r.t.\ our new fit (symbols) and
the DSS analysis (dashed lines).
\label{fig:ee-untagged}}
%
\vspace*{-0.5cm}
\begin{center}
\epsfig{figure=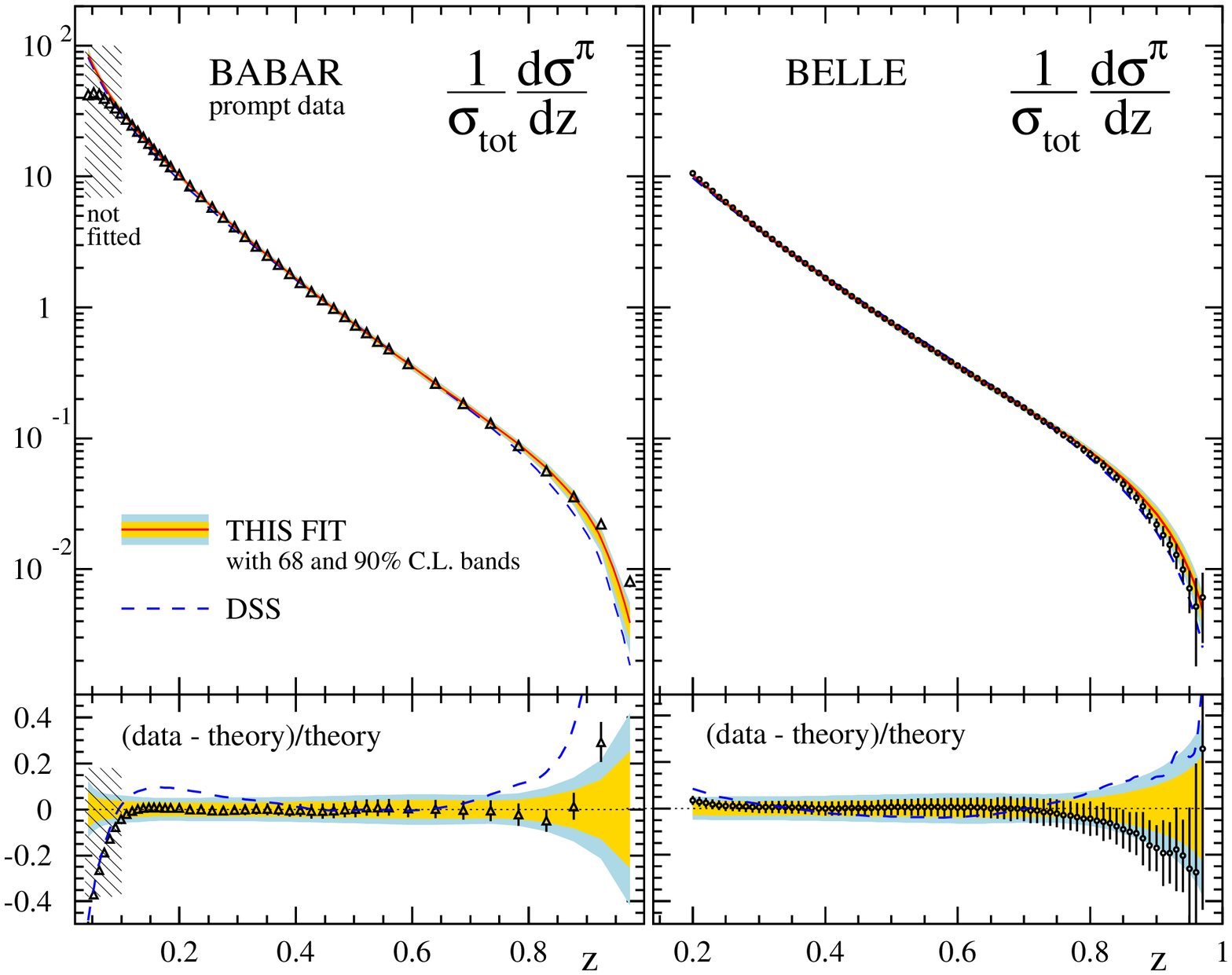,width=0.62\textwidth}
\end{center}
\vspace*{-0.7cm}
\caption{Left-hand side: comparison of our new NLO results (solid line) 
with the new {\sc BaBar} ``prompt'' data \cite{ref:babardata}; also
shown is the result obtained with the DSS fit \cite{ref:dss} (dashed line).
Right-hand side: same, but now for the Belle data \cite{ref:belledata}.
The lower panels show ``(data-theory)/theory'' for each of the data sets w.r.t.\ our new fit (symbols) and
the DSS analysis (dashed lines).
The inner and outer shaded bands correspond to the new uncertainty estimates at $68\%$ and $90\%$ C.L.,
respectively.
\label{fig:ee-babarbelle}}
\end{figure*} 
The {\sc Belle} data \cite{ref:belledata}, shown in the right-hand panel of Fig.~\ref{fig:ee-babarbelle},
provide not only the finest binning in $z$ but also reach the
highest $z$ values measured so far. Above $z\gtrsim 0.8$ one observes an increasing
trend for the new fit to overshoot the data, but still within the estimated and growing theoretical uncertainties though.
In this kinematic regime one expects large logarithmic corrections, which appear
in each order of perturbation theory, to become more and more relevant. 
It is known how to resum such terms to all orders in the strong coupling \cite{ref:resum}, and
it might be worthwhile to explore their relevance in a future dedicated analysis 
and whether they could further improve the agreement with data. Resummations also provide 
a window to non-perturbative contributions to the perturbative series so far little explored.
The binning of {\sc BaBar} data \cite{ref:babardata} is more sparse towards large $z$, 
and a similar trend as for the {\sc Belle} data is not visible here.

For all the sets shown in Fig.~\ref{fig:ee-untagged}, the new fit
is able to follow the trend of the data even below the $z$ values
included in the analysis (the region indicated by the hatched area).
Agreement with {\sc BaBar} data below the cut $z=0.1$ quickly deteriorates though.
In this region, the data start to drop while the NLO SIA cross section continues to rise
as can be seen in the left-hand panel of Fig.~\ref{fig:ee-babarbelle}.
Since the {\sc BaBar} data are taken at the lowest c.m.s.\ energy, such an effect is
not unexpected and signifies the onset of neglected hadron mass effects in the theoretical
framework. In fact, this was the reason for us to choose a somewhat higher cut in $z$, $z>0.1$,
than for the other SIA data obtained at higher c.m.s.\ energies.
The {\sc Belle} experiment did not publish any data below $z=0.2$ \cite{ref:belledata}.

\begin{figure*}[tbh!]
\vspace*{-0.4cm}
\begin{center}
\epsfig{figure=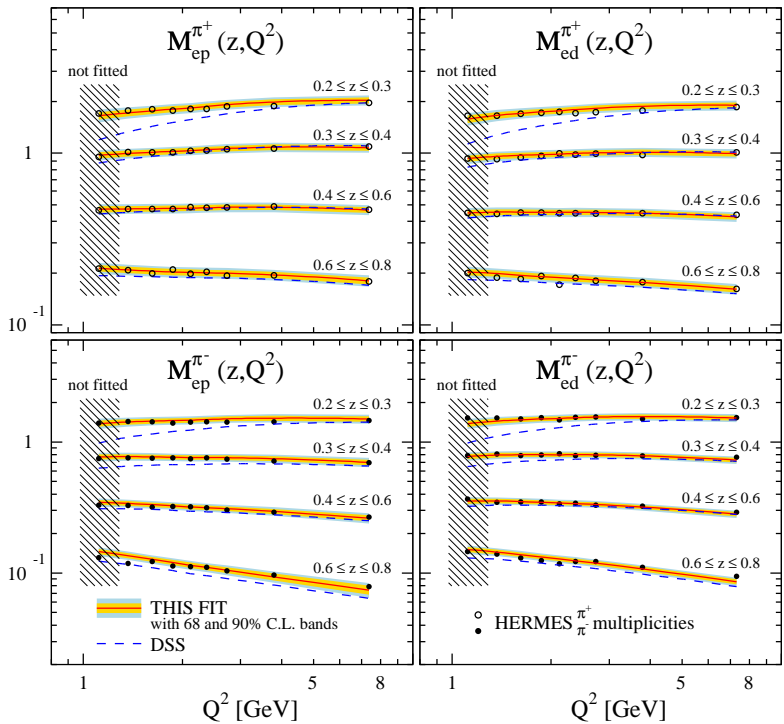,width=0.65\textwidth}
\end{center}
\vspace*{-0.5cm}
\caption{Comparison of our NLO results for charged pion multiplicities in SIDIS 
off proton (left panels) and deuteron (right panels) targets with data
from the {\sc{Hermes}} Collaboration \cite{ref:hermesmult}.
The inner and outer shaded bands correspond to uncertainty estimates at $68\%$ and $90\%$ C.L.,
respectively. Also shown are the results obtained with the DSS FFs (dashed lines).
\label{fig:sidis-hermes}}
\end{figure*} 
Also shown in Figs.~\ref{fig:ee-untagged} and \ref{fig:ee-babarbelle} are the theoretical
results obtained with the original DSS FFs (dashed lines), i.e., without any refitting or
adjusting normalization shifts. The agreement with SIA data is
in general very good, except for some small deviations from the recent $B$ factory data,
most noticeable in the comparison to {\sc BaBar}.
Contrary to the new analysis, the original DSS fit 
undershoots both the {\sc Belle} and {\sc BaBar} data at high $z$.

Our estimated uncertainty bands,
also shown in Figs.~\ref{fig:ee-untagged} and \ref{fig:ee-babarbelle},  
reflect the accuracy and kinematical coverage of the fitted data. 
They increase towards both small and large $z$, similar to the
pattern observed for the individual $D_i^{\pi^{+}}$ 
in Figs.~\ref{fig:ff-at-10} and \ref{fig:ff-at-mz}.
One should keep in mind that the obtained bands are constrained by the fit 
to the global set of SIA, SIDIS, and $pp$ data and do not necessarily 
have to follow the accuracy of each individual set of data.

As was already mentioned in Sec.~\ref{sec:ff-results}, the SIA data from the LEP and SLAC experiments
constrain mainly the total quark singlet fragmentation to pions as
up-type and down-type quark couplings to the exchanged $Z$ gauge boson are roughly 
equal at $Q\simeq M_Z$. The new {\sc BaBar} and {\sc Belle} data are dominated by photon exchange
and, hence, prefer up-type quark flavors. When combined, this leads to some partial flavor separation.
QCD scale evolution between $Q^2\simeq 110\,\mathrm{GeV}^2$ and $Q^2=M_Z^2$
provides some additional constraints, in particular, also for the gluon FF. 
The flavor-tagged LEP and SLAC data, listed in Tab.~\ref{tab:exppiontab}, 
are still the best ``direct'' source of information on the charm- and bottom-to-pion FFs.

Finally, we wish to remark that despite the excellent agreement with all SIA data there are still
some issues which require further scrutiny and, perhaps, more detailed comparisons among the
different experimental groups. 
One concern is the question to what extent ``feed-down'' pions from weak decays 
contribute to the individual data sets. Different treatments of QED radiative corrections,
whose main effect is to lower the ``true'' c.m.s.\ energy $\sqrt{S}$ of the collisions,
might be another source of potential tension. For instance, the {\sc Belle} Collaboration \cite{ref:belledata}
provides only a measurement of the cross section $d\sigma/dz$, while all other
experiments in SIA scale their quoted results by the total cross section $\sigma_{\mathrm{tot}}$
for $e^+e^- \to \mathrm{hadrons}$.
Since {\sc Belle} cuts on radiative photon events if their energy exceeds a certain threshold,
rather than attempting to unfold the radiative QED effects, one has to take this into account
when normalizing the {\sc Belle} data to the conventional 
$1/\sigma_{\mathrm{tot}}\; d\sigma/dz$ in a global fit.

\begin{figure*}[th!]
\vspace*{-0.5cm}
\begin{center}
\epsfig{figure=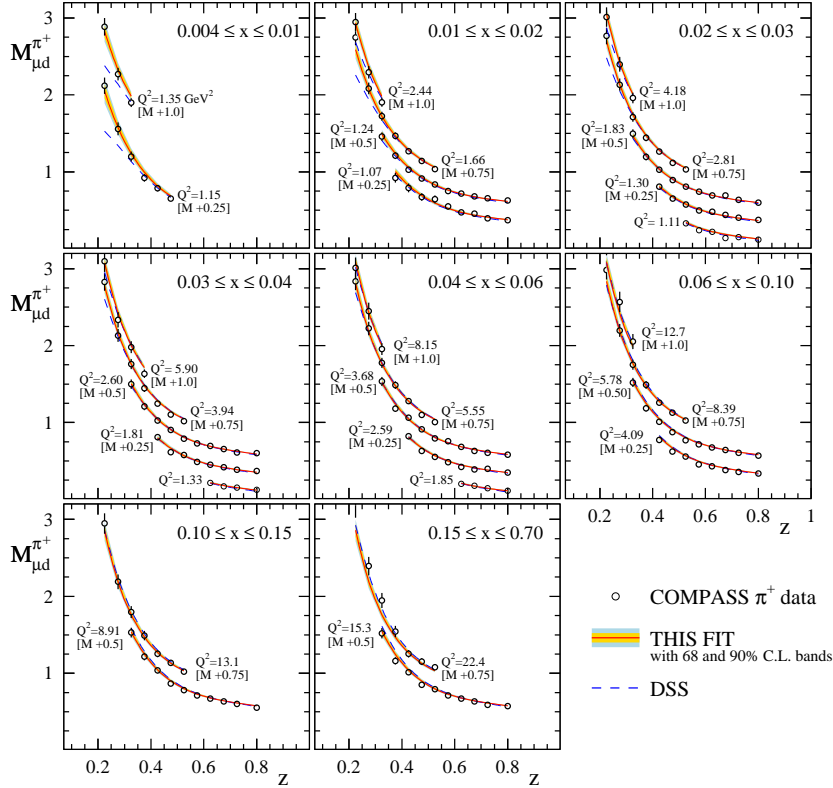,width=0.66\textwidth}
\end{center}
\vspace*{-0.75cm}
\caption{Comparison of our NLO results for $\pi^+$ multiplicities in SIDIS of muons off a
deuteron target with preliminary data
from the {\sc{Compass}} experiment \cite{ref:compassmult} for various bins in $x$ and $Q^2$.
The inner and outer shaded bands correspond to our uncertainty estimates at $68\%$ and $90\%$ C.L.,
respectively. Also shown are the results obtained with the DSS FFs (dashed lines).
As indicated in the plot, different constant factors are added to the multiplicities $M_{\mu d}^{\pi^{+}}$ to distinguish
results for different values of $Q^2$ in the same $x$ bin.
\label{fig:sidis-compass-piplus}}
\end{figure*} 
%
\subsection{Semi-Inclusive DIS Multiplicities \label{sec:sidis-data}}
%
The most powerful constraint of flavor-separated FFs comes from charged pion
multiplicities in SIDIS. Contrary to SIA, which produces $\pi^+$ and $\pi^-$ at
equal rates, multiplicities are sensitive to the produced hadron's charge
through the choice of the target hadron in DIS. 
For instance, data taken on a proton target will produce more $\pi^+$ than $\pi^-$,
since $u$-quarks are more abundant in a proton than $d$-quarks, and they 
are also preferred in their coupling to the probing virtual photon 
due to their larger electrical charge.

Compared to the DSS analysis, where we only had some preliminary set of
pion multiplicities on a deuteron target from the {\sc Hermes} Collaboration 
at our disposal \cite{ref:hermes-old}, we can now use their recently published, final set of data
for both proton and deuteron targets \cite{ref:hermesmult}. 
In Fig.~\ref{fig:sidis-hermes} we illustrate the quality of the new fit with respect to
the {\sc Hermes} data. Shown are the charged pion multiplicities $M_{e,p(d)}^{\pi^{\pm}}$, which are defined as
the ratio of the inclusive pion yield and the total DIS cross section at the same $x$ and $Q^2$ values (bins)
in electron-proton ($ep$) or electron-deuteron ($ed$) scattering:
\begin{equation}
\label{eq:mult}
M_{e,p(d)}^{\pi^{\pm}} \equiv \frac{ d\sigma^{\pi^{\pm}}/dx\,dQ^2\,dz}
{d\sigma/dx\,dQ^2}\;\;.
\end{equation}
The extraction of the FFs requires knowledge of the PDFs of the proton (deuteron) target
for which we use the NLO parametrization of the MSTW Collaboration \cite{ref:mstw}
as was already mentioned above.
In the fit we consider the projection of the three-dimensional multiplicity data 
onto the $Q^2$ dependence for four different bins of the pion's momentum fraction $z$, 
which is most sensitive to the quantities we are interested in, the parton-to-pion FFs. 
The $x$ integrated ratio (\ref{eq:mult}) is also least sensitive to the actual choice of PDFs.
We use the standard Mellin technique \cite{ref:mellin2} to pre-calculate look-up tables for each data point 
at NLO accuracy to speed up the fitting procedure and to facilitate the uncertainty analyses
significantly.
We recall that at NLO, the relevant hard scattering coefficient functions for SIDIS \cite{ref:sidis-nlo,ref:lambda-nlo}
depend in a non-trivial way on both $x$ and $z$, such that an often used naive approximation,
where the $x$ and the $z$ dependence in Eq.~(\ref{eq:mult}) is assumed to completely factorize, is bound 
to fail. Even at LO accuracy such an assumption cannot work as soon as different quark
flavors fragment differently into the observed hadron which they do for charged pions.

\begin{figure*}[ht]
\vspace*{-0.5cm}
\begin{center}
\epsfig{figure=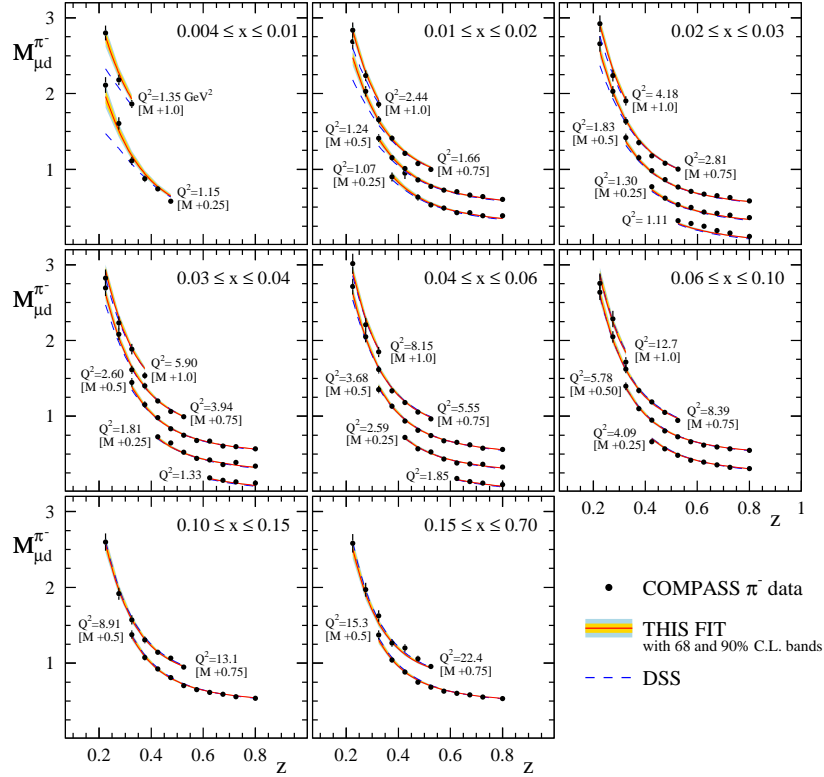,width=0.65\textwidth}
\end{center}
\vspace*{-0.75cm}
\caption{As in Fig.~\ref{fig:sidis-compass-piplus} but now for $\pi^-$ multiplicities.
\label{fig:sidis-compass-piminus}}
\end{figure*} 
The agreement between the {\sc Hermes} data \cite{ref:hermesmult} and the updated fit is remarkably better than with
the preliminary results \cite{ref:hermes-old} used in the DSS analysis, see Fig.~4 in Ref.~\cite{ref:dss}. 
This is largely due to the much improved
precision of the final data, which also exhibit considerably less fluctuations from bin to bin, 
in particular, for $\pi^-$. 
This is also reflected in the total $\chi^2$ for the {\sc Hermes} data set, which reduces from 188.2 for
64 data points on a proton target in the DSS fit \cite{ref:dss} to 175 for 128 data points in the current analysis.
In Fig.~\ref{fig:sidis-hermes} we compare again also to the result of a calculation based on the
DSS FFs, without any re-fitting or adjusting normalizations. As can be seen, the agreement
with data is not optimal, and the theory predictions fall short of the data in all bins.
Most noticeable is the disagreement at $0.2\le z \le 0.3$ and $0.6\le z \le 0.8$ for both
$\pi^+$ and $\pi^-$ data. Here, the DSS result is well outside our current uncertainty estimates
shown, as before, as shaded bands in Fig.~\ref{fig:sidis-hermes}.

The use of the {\sc Hermes} multiplicity data as a means of providing a reliable flavor and charge separation
for pion FFs in the DSS fit was often questioned in the past because of the smallish $Q^2$ values 
of some of the data points.
New, still preliminary data from the {\sc Compass} Collaboration \cite{ref:compassmult}, taken at a 
higher c.m.s.\ energy, will shed some light on the validity of using a standard, leading-twist 
pQCD framework at NLO accuracy \cite{ref:sidis-nlo,ref:lambda-nlo} to describe the {\sc Hermes}  multiplicity data
for charged pions.

In the present fit we can use charged pion results from {\sc Compass} obtained on a deuteron target \cite{ref:compassmult}.
More specifically, the data are presented as a function of $z$ in 8 bins of $x$, each
subdivided into various bins in $Q^2$. In total 199 data points pass our cuts for
both $\pi^+$ and $\pi^-$. The comparison of the {\sc Compass} data to the results of our fit is
presented in Figs.~\ref{fig:sidis-compass-piplus} - \ref{fig:sidis-compass-dmt}.
A very satisfactory agreement is achieved in almost all bins across the entire kinematic regime
covered by data, as can be best inferred from Fig.~\ref{fig:sidis-compass-dmt}, 
where we show ``(data-theory)/theory''. The obtained $\chi^2/{\mathrm{d.o.f.}}$ for both $\pi^+$ and $\pi^-$ multiplicities
is close to unity, see Tab.~\ref{tab:exppiontab},
demonstrating that the low energy {\sc Hermes} \cite{ref:hermesmult} and the 
{\sc Compass} \cite{ref:compassmult} data
can be described simultaneously and without spoiling the agreement with SIA results.
For comparison we show again theoretical results obtained with the
DSS FFs (dashed lines), which also agree well with {\sc Compass} data except for some of the bins corresponding
to the lowest $Q^2$ values. This is in line with the observations for the {\sc Hermes} data above,
where the deviations with DSS were found to be largest at the smallest $Q^2$.
%
\begin{figure*}[ht]
\vspace*{-0.5cm}
\begin{center}
\epsfig{figure=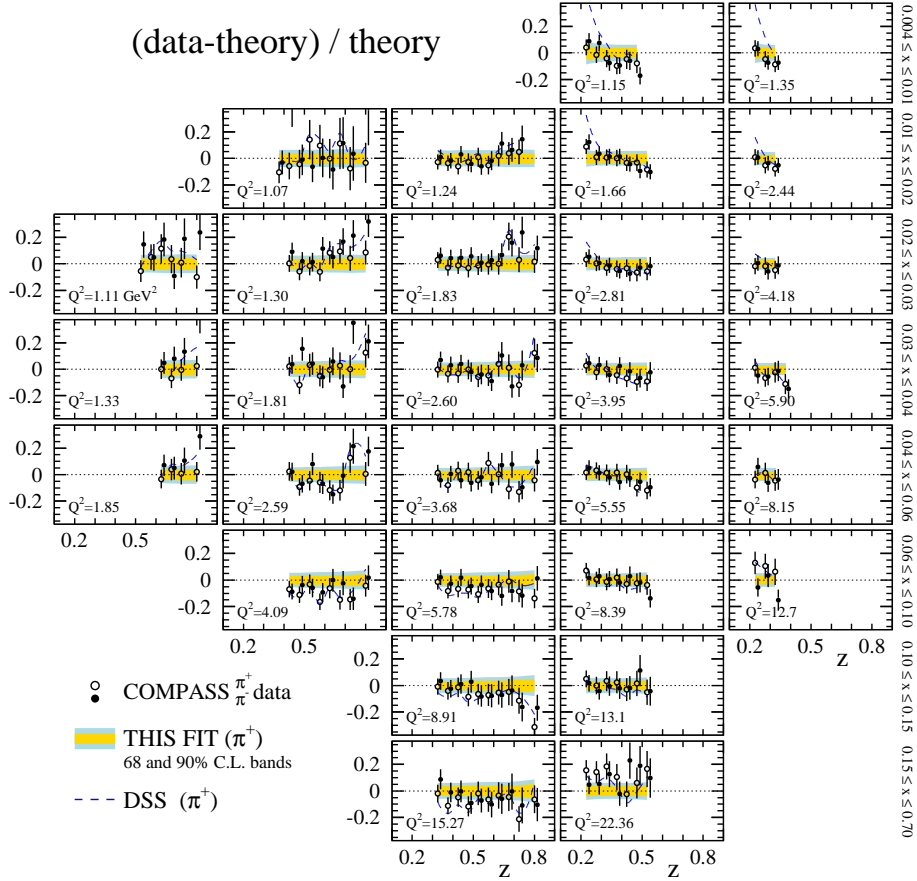,width=0.74\textwidth}
\end{center}
\vspace*{-0.75cm}
\caption{As in Figs.~\ref{fig:sidis-compass-piplus} and \ref{fig:sidis-compass-piminus}
but now showing ``(data-theory)/theory'' for our new NLO fit (open and closed circles correspond
to $\pi^+$ and $\pi^-$ multiplicities, respectively) in each $x$ and $Q^2$ bin. 
The shaded uncertainty bands and the results
obtained with the DSS FFs (dashed lines) are for $\pi^+$ production.
\label{fig:sidis-compass-dmt}}
\end{figure*} 

As was already mentioned in Sec.~\ref{sec:ff-results}, the new SIDIS data now favor
almost identical $u+\bar{u}$ and $d+\bar{d}$ FFs, i.e., very little or no charge symmetry breaking.
This is also preferred by data on the $\pi^-/\pi^+$ ratio in $pp$ collisions which we
discuss next. 

\subsection{RHIC and LHC Data \label{sec:pp-results}}
%
The last of the three pillars of our global analysis of parton-to-pion FFs
is the wealth of experimental information coming from hadron-hadron collisions,
more specifically, single-inclusive high-$p_T$ pion production in $pp$ collisions
at BNL-RHIC and CERN-LHC.
Compared to the original DSS analysis \cite{ref:dss}, which mainly made use of the {\sc Phenix}
data for $\pi^0$ production at mid rapidity \cite{ref:phenixdata}, we now have, in addition, results
from the {\sc Star} Collaboration for neutral and charged pions 
\cite{ref:starcharged06,ref:stardata09,ref:starratio11,ref:stardata13} as well as
first data from the LHC \cite{ref:alicedata}. 

Due to the complexity of the underlying hard-scattering processes at NLO accuracy \cite{ref:pp-nlo},
the use of a fast, grid-based method such as the Mellin technique, 
to implement the relevant expressions efficiently and without the need of any approximations
is indispensable here. As in various previous analyses \cite{ref:dss,ref:dss2,ref:eta,ref:dssv}, and for the implementation
of the SIDIS multiplicities in NLO, we adopt the well-tested method based on Mellin moments
as described in Ref.~\cite{ref:mellin2}.
Since inclusive particle spectra at not too large values of $p_T$ are dominated
by gluon-induced processes in $pp$ collisions \cite{ref:us-lhc}, the RHIC and LHC data will provide
invaluable information on the otherwise only weakly constrained gluon
FF $D_g^{\pi^{+}}$. 
  
One of the main results of our updated fits is to reveal a tension between the $p_T$ spectra of
neutral pions measured at the RHIC experiments and by the {\sc Alice} Collaboration.
In some sense this was already anticipated by comparisons to expectations obtained with
the previous DSS FFs, 
which are known to describe the RHIC data nicely down to $p_T\simeq 1.5\,\mathrm{GeV}$ \cite{ref:phenixdata,ref:dss}
but were found to grossly overshoot recent {\sc Alice} results at $\sqrt{s}=7\,\mathrm{TeV}$ for
essentially all $p_T$ values \cite{ref:alicedata,ref:helenius}.
We have tried to accommodate both sets of $pp$ data together by introducing additional freedom
to our standard functional form in Eq.~(\ref{eq:ff-input}) but to no avail.
In particular, at smallish $p_T$ values, below about $5\,\mathrm{GeV}$, the two sets of data 
appear to be mutually exclusive in a global fit.

\begin{figure}[ht!]
\vspace*{-0.5cm}
\begin{center}
\epsfig{figure=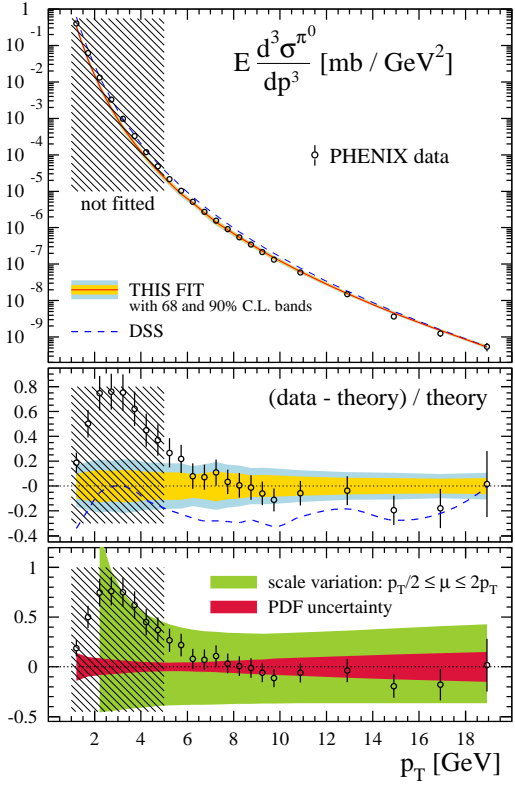,width=0.45\textwidth}
\end{center}
\vspace*{-0.5cm}
\caption{Comparison of our NLO results for single-inclusive neutral pion production in
$pp$ collisions at $\sqrt{S}=200\,\mathrm{GeV}$ with the {\sc{Phenix}} data \cite{ref:phenixdata} (upper panel).
The inner and outer shaded bands correspond to uncertainty estimates at $68\%$ and $90\%$ C.L.,
respectively. Also shown are the results obtained with the DSS FFs (dashed line).
The middle panel shows corresponding results for ``(data-theory)/theory''.
In the lower panel we illustrate the relevance of theoretical uncertainties due to scale 
and PDF variations.
\label{fig:pp-phenix}}
\end{figure} 
Since we do not want to remove either of the data sets from the analysis
and, in any case, have no means of judging whether there is a potential 
experimental inconsistency among the different $pp$ sets, we decided to introduce
a cut on the $p_T$ of the observed pion. Including only $pp$ data with
$p_T \ge 5\,\mathrm{GeV}$ largely resolves the observed tension between RHIC
and LHC data.
This is illustrated in Figs.~\ref{fig:pp-phenix} and \ref{fig:pp-alice}
where we compare to the {\sc Phenix} \cite{ref:phenixdata} and {\sc Alice} \cite{ref:alicedata} data, respectively.
We note that the calculated normalization shift (\ref{eq:normshift}) for 
the {\sc Alice} $7\,\mathrm{TeV}$ data results in a down-shift, outside
the experimentally quoted normalization uncertainty, which contributes
significantly to the quoted $\chi^2$ value in Tab.~\ref{tab:exppiontab}.
As we have already hinted at in our discussion of the SIA data in Sec.~\ref{sec:sia-data}, a different
treatment of decay pions by the RHIC and LHC experiments might play some
role for the tension observed at $p_T\lesssim 5\, \mathrm{GeV}$.

\begin{figure}[ht!]
\vspace*{-0.5cm}
\begin{center}
\epsfig{figure=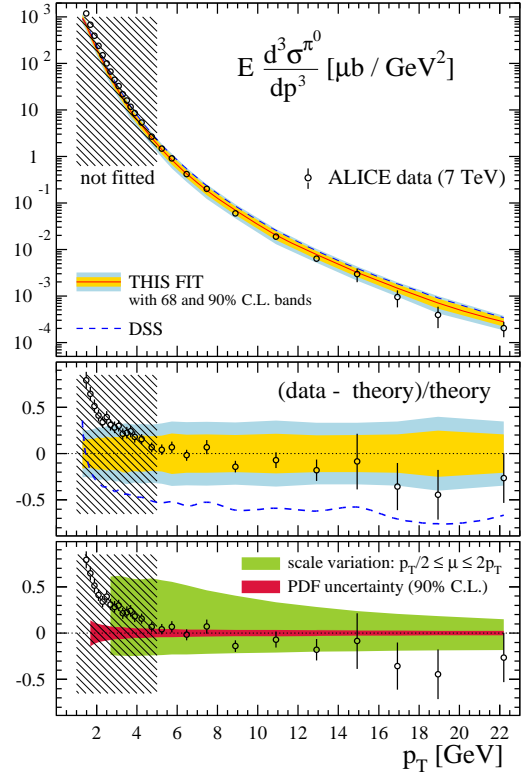,width=0.45\textwidth}
\end{center}
\vspace*{-0.5cm}
\caption{As in Fig.~\ref{fig:pp-phenix} but now for $\pi^0$ production in $pp$ collisions
at $\sqrt{S}=7\,\mathrm{TeV}$ as measured by the {\sc{Alice}} experiment \cite{ref:alicedata}.
\label{fig:pp-alice}}
\end{figure} 
As can be seen, both data sets are well described by the global fit above
the introduced $p_T$ cut which is indicated by the hatched area in both figures.
One also notices the still sizable theoretical scale ambiguity at NLO accuracy,
which is indicated in the lower panels of Figs.~\ref{fig:pp-phenix} and \ref{fig:pp-alice}
and within which the data are consistent with the fit even below the imposed $p_T$ cut.
The PDF uncertainties, computed with the $90\%$ CL NLO sets from MSTW \cite{ref:mstw} and also illustrated
in the same panels, are much less significant than the scale ambiguities, in particular,
for the {\sc Alice} data.

The resulting global fit is, as always, a compromise of all the data sets included in the analysis and,
in particular, mediates between RHIC $pp$ data preferring a larger gluon-to-pion FF and
LHC {\sc Alice} data favoring a smaller $D_g^{\pi^{+}}$. The net effect is a significantly
reduced $D_g^{\pi^{+}}$ as compared to the DSS fit \cite{ref:dss}, as was already discussed in
Sec.~\ref{sec:ff-results} and illustrated in Fig.~\ref{fig:ff-at-10}.
Because of the remaining small tension, the estimated uncertainties on $D_g^{\pi^{+}}$
are sizable, despite the available amount of rather precise experimental data from $pp$ collisions. 
If both RHIC and LHC data would point to a more similar $D_g^{\pi^{+}}$, the resulting uncertainties 
would likely to be somewhat smaller, however, the large theoretical scale ambiguities illustrated
above still remain. We note that $pp$ data at mid rapidity dominantly probe the gluon
FF at medium-to-large $z$ values as was, for instance, demonstrated in Ref.~\cite{ref:us-lhc}

\begin{figure}[ht!]
\vspace*{-0.5cm}
\begin{center}
\epsfig{figure=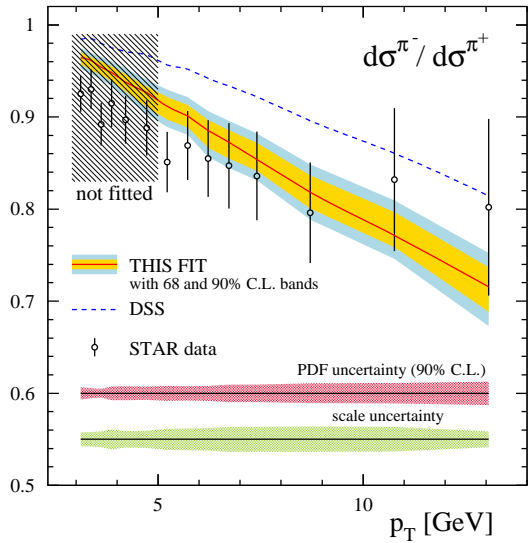,width=0.465\textwidth}
\end{center}
\vspace*{-0.5cm}
\caption{Comparison of our NLO results for the $\pi^-/\pi^+$ cross section ratio in
$pp$ collisions at $\sqrt{S}=200\,\mathrm{GeV}$ with the {\sc{Star}} data \cite{ref:starratio11}.
The inner and outer shaded bands correspond to uncertainty estimates at $68\%$ and $90\%$ C.L.,
respectively. Also shown are the results obtained with the DSS FFs (dashed line).
Scale and PDF uncertainties are indicated at the base of the plot.
\label{fig:pp-pi-ratio-star}}
\end{figure} 
The last two figures give a similar comparison to the {\sc Star} data 
\cite{ref:starcharged06,ref:stardata09,ref:starratio11,ref:stardata13} for which we
adopt, of course, the same $p_T$ cut as for the other $pp$ sets. None of these results
was included in the DSS analysis.
In Fig.~\ref{fig:pp-pi-ratio-star} we focus on the $\pi^-/\pi^+$ ratio at mid rapidity
\cite{ref:starratio11}, which is now much better described by the fit than with the DSS FFs.
Scale ambiguities partially cancel in the ratio and are much less dramatic than
for the individual cross sections, cf.~Fig.~\ref{fig:pp-phenix}.
As was already mentioned, the ratio is sensitive to a potential charge asymmetry or
SU(2) breaking, as parametrized by Eq.~(\ref{eq:su2breaking}) in out fit. Like for the SIDIS multiplicities,
the fit prefers little or no breaking, i.e., $N_{d+\bar{d}}$ in (\ref{eq:su2breaking})
close to unity.

\begin{figure}[th!]
\vspace*{-0.5cm}
\begin{center}
\epsfig{figure=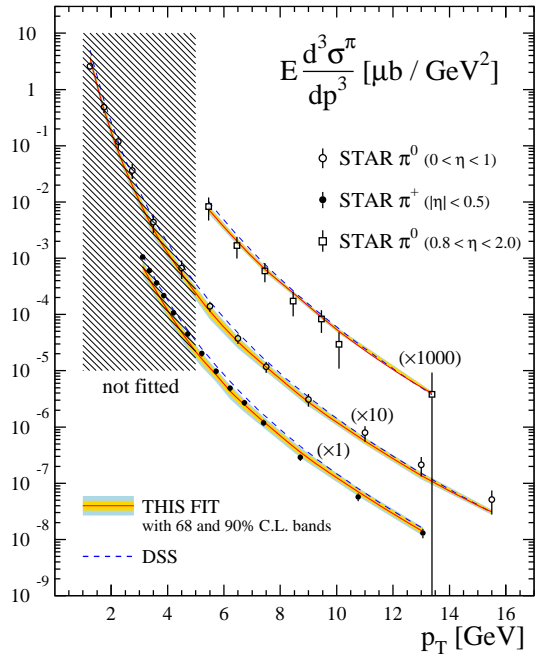,width=0.465\textwidth}
\end{center}
\vspace*{-0.5cm}
\caption{Comparison of our NLO results for single-inclusive $\pi^0$ and $\pi^+$ production
in different rapidity ranges in
$pp$ collisions at $\sqrt{S}=200\,\mathrm{GeV}$ with the corresponding {\sc{Star}} data
\cite{ref:starcharged06,ref:stardata09,ref:stardata13}.
The inner and outer shaded bands correspond to uncertainty estimates at $68\%$ and $90\%$ C.L.,
respectively. Also shown are the results obtained with the DSS FFs (dashed lines).
Note that some of the results are rescaled for clarity.
\label{fig:pp-star-3in1}}
\end{figure}
Figure~\ref{fig:pp-star-3in1} gives on overview of the three other sets of 
single-inclusive pion data from the {\sc Star} Collaboration used in the fit
\cite{ref:starcharged06,ref:stardata09,ref:stardata13}, which span different rapidity intervals. 
Since we fit to the $\pi^-/\pi^+$ ratio shown in Fig.~\ref{fig:pp-pi-ratio-star}
and $\pi^+$ data, we exclude results on the $\pi^-$ cross section to avoid 
double-counting. The description of the data is very good, even below the $p_T$-cut
of $5\,\mathrm{GeV}$, indicating that there is a little bit less of a tension with
{\sc Alice} results than for the {\sc Phenix} experiment. Calculations based on
the DSS FFs (dashed lines) also provide a good description of data.

\section{Summary and Outlook}
%
We have presented a new, comprehensive global QCD analysis of parton-to-pion fragmentation functions 
at next-to-leading order accuracy including the latest experimental information.
The analyzed data for inclusive pion production in semi-inclusive electron-positron
annihilation, deep-inelastic scattering, and proton-proton collisions span
energy scales ranging from about $1\,\mathrm{GeV}$ up to the mass of the $Z$ boson.
The achieved, very satisfactory and simultaneous description of all data sets strongly supports the 
validity of the underlying theoretical framework based on pQCD and, in particular, the notion of
factorization and universality for parton-to-pion fragmentation functions.

Compared to our previous analysis, which was based on much less precise experimental input
and to which we have made extensive comparisons throughout this work,
we now obtained a significantly better fit, as measured in terms its the global $\chi^2$, 
using the same functional form with only a few additional fit parameters. 
While most of the favored and unfavored quark-to-pion fragmentation functions are by and large
similar to our previous results, the reduced amount of pions stemming for the hadronization
of gluons is a noteworthy outcome of the new analysis.
This finding was driven by first data from the CERN-LHC experiments, which, surprisingly, turned out
to be mutually incompatible with previously available data obtained in lower center-of-mass system
energy collisions at BNL-RHIC. 
To remedy this tension in our fit, we were forced to introduce a lower cut on the transverse momentum
of the produced pions in proton-proton collisions.
We have argued that it should be worthwhile for the experiments to compare in detail their procedures to
determine pion yields as, for instance, different cuts for secondary pions from decays of other, heavier mesons
perhaps have some numerical impact.
We believe that such a contamination from feed-down pions might show up most prominently 
at small transverse momenta, where we currently observe the tension between RHIC and LHC data.
We wish to mention that in the quark sector, the new data do not favor any charge symmetry violation 
between the total up- and down-quark fragmentation functions, contrary to our previous fit.

We have also performed a, what we believe, first reliable and trustworthy estimate of uncertainties for
parton-to-pion fragmentation functions based on the standard iterative Hessian method. This was made possible by
the wealth of new data included in our updated global analysis.
The obtained uncertainties are still sizable and range at best from about ten percent to twenty-five
percent for the total $u$-quark and gluon fragmentation function, respectively, in the kinematic
regions covered by data and they quickly deteriorate beyond.
A new asset of the current analysis is the analytic procedure to determine the optimum normalization
shift for each data set in the fit, which greatly facilitated the global fitting procedure

The newly obtained pion fragmentation functions and their uncertainty estimates will be a crucial 
ingredient in future global analyses of both helicity and transverse-momentum dependent parton
densities, which heavily draw on data with identified pions in the final-state.
Our results will also serve as the baseline in heavy ion and proton-heavy ion collisions,
where one of the main objectives is to quantify and understand possible modifications of
hadron production yields by the nuclear medium.
Also, the current analysis framework will be adopted for updates of the parton-to-kaon
fragmentation functions, which we will 
pursue once all the promised sets of new data eventually become available. 
Since pions and kaons constitute by far the largest fraction in frequently measured 
yields of unidentified charged hadrons, 
a precise determination of their respective, optimum sets of fragmentation functions,
including reliable uncertainty estimates, is critical to determine the 
room left for other hadrons, such as protons, in a future global analysis of charged hadron data.

Further improvements of parton-to-pion fragmentation functions from the theory side
should include an improved treatment of heavy quark-to-pion fragmentation functions,
likely along similar lines as for heavy flavor parton densities. Also, the impact
of higher order corrections beyond the next-to-leading order accuracy 
should be explored. On the one hand, it is already possible to
perform a next-to-next-to-leading order analysis of electron-positron annihilation data,
and, on the other hand, the theoretical framework for all-order resummations of potentially
large logarithmic corrections is available.
On the experimental side,
RHIC and the LHC will continue to provide new data on identified hadron spectra but it
should be also worthwhile to explore the potential of future accelerator projects, such as
an Electron-Ion Collider currently pursued in the U.S. \cite{ref:eic}, to further our
knowledge of fragmentation functions and the physics behind hadronization.

\section*{Acknowledgments}
%
We are grateful to M.\ Leitgab and R.\ Seidl ({\sc Belle}) and 
F.\ Kunne, N.\ Makke, and R.\ Windmolders ({\sc Compass})
for helpful discussions about their measurements.
This work was supported in part by CONICET, ANPCyT, UBACyT, CONACyT-Mexico, 
the Research Executive Agency (REA) of the European Union under the Grant 
Agreement number PITN-GA-2010-264564 (LHCPhenoNet)
and the Institutional Strategy of the University of T\"{u}bingen (DFG, ZUK 63).  


%
\end{document}